\RequirePackage{amsmath}
\documentclass[runningheads]{llncs}
\usepackage[utf8]{inputenc}
\usepackage[misc]{ifsym}
\usepackage{amssymb}
\usepackage{xcolor}
\usepackage{graphicx}
\usepackage{babel}
\usepackage{multicol}
\usepackage{soul}
\usepackage{multirow}
\usepackage{enumitem}

\usepackage{hyperref}

\usepackage{microtype}

\usepackage[caption=false]{subfig}

\usepackage{tikz}
\usetikzlibrary{arrows,shapes.geometric,shapes.misc,decorations,automata,backgrounds,petri,positioning,patterns}
\usetikzlibrary{}

\usepackage{wrapfig}

\usepackage{siunitx}
\newcolumntype{?}{!{\vrule width 1.5pt}}
\usepackage{array}

\makeatletter
\newcommand{\thickhline}{%
    \noalign {\ifnum 0=`}\fi \hrule height 1.5pt
    \futurelet \reserved@a \@xhline
}

\newcommand{\sst}[1]{\mbox{\scriptsize #1}}

\newcommand{\bg}{{\bf{g}}}
\newcommand{\bv}{{\bf{v}}}

\newcommand{\bG}{{\bf{G}}}
\newcommand{\bA}{{\bf{A}}}
\newcommand{\ba}{{\bf{a}}}
\newcommand{\bB}{{\bf{B}}}
\newcommand{\bI}{{\bf{I}}}
\newcommand{\by}{{\bf{y}}}

\newcommand{\bU}{{\bf{U}}}
\newcommand{\bs}{{\bf{s}}}
\newcommand{\bq}{{\bf{q}}}
\newcommand{\bx}{{\bf{x}}}
\newcommand{\bw}{{\bf{w}}}

\newcommand{\bOne}{{\boldsymbol{1}}}
\newcommand{\bdelta}{\boldsymbol{\delta}}

\begin{document}

\title{Learning Optimal Linear Precoding for Cell-Free Massive MIMO with GNN}

\author{Benjamin Parlier\inst{1} \and
Lou Sala\"un\inst{2} (\Letter) \and
Hong Yang\inst{3}}

\authorrunning{B. Parlier et al.}

\institute{Université Paris-Saclay, Centrale-Supélec, Gif-sur-Yvette, 91190 France
\and
Nokia Bell Labs, Massy, 91300 France \email{lou.salaun@nokia-bell-labs.com}
\and
Nokia Bell Labs, Murray Hill, NJ 07974 USA}

\maketitle

\begin{abstract}
We develop a graph neural network (GNN) to compute, within a time budget of 1 to 2 milliseconds required by practical systems, the optimal linear precoder (OLP) maximizing the minimal downlink user data rate for a Cell-Free Massive MIMO system – a key 6G wireless technology.
The state-of-the-art method is a bisection search on second order cone programming feasibility test (B-SOCP) which is a magnitude too slow for practical systems.  
Our approach relies on representing OLP as a node-level prediction task on a graph. We construct a graph that accurately captures the interdependence relation between access points (APs) and user equipments (UEs), and the permutation equivariance of the Max-Min problem.
Our neural network, named OLP-GNN, is trained on data obtained by B-SOCP.
We tailor the OLP-GNN size, together with several artful data preprocessing and postprocessing methods to meet the runtime requirement.
We show by extensive simulations that it achieves near optimal spectral efficiency in a range of scenarios with different number of APs and UEs, and for both line-of-sight
and non-line-of-sight radio propagation environments.
\end{abstract}

\section{Introduction}
We employ a graph neural network (GNN) to solve an important problem relating to a key 6G wireless technology – Cell-Free Massive MIMO (CFmMIMO). The concept of CFmMIMO was first introduced in \cite{ym2013} and further analyzed in~\cite{elina2017,ngo2017}. “MIMO” refers to “Multiple Input Multiple Output” that takes advantage of spatial multiplexing to serve multiple users simultaneously, thereby greatly increases the spectral efficiency in terms of bits per second per Hertz. “Massive” refers to the hundreds of service antennas in the systems. “Cell-Free”, in contrast to cellular, refers to a wireless network where a large number of access points (APs) are distributed in a geographic area to jointly serve a collection of users simultaneously. CFmMIMO relies on a well-designed precoder to beamform ultrahigh data rate to users. We construct a GNN to compute, within a time budget of 1 to 2 milliseconds, the optimal linear precoder (OLP) that maximizes the minimal (Max-Min) downlink user data rate for a CFmMIMO.

Motivations for Max-Min, OLP and time budget are summarized in the following.
Max-Min is highly desirable because all wireless systems aim to achieve the highest data rate possible for all users. A key advantage of CFmMIMO is all the users in the system have statistically identical large scale fading profiles. This contrasts with the traditional cellular system in which large scale fading profiles are uneven for users near the base stations and users at the cell edge. Conceptually, achieving equal throughputs for all users is natural for CFmMIMO. 
OLP is the optimal precoder among all linear precoders. Furthermore, it is effectively the optimal precoder for massive MIMO. By virtue of the law of large numbers, many service antennas effectively orthogonalize the communication channels, thereby making linear precoding substantially optimal \cite{marzetta2016fundamentals}. 
Millisecond scale time budget is critical for real world applications. For typical mobility applications in urban and suburban scenarios \cite{ym_greencom}, a new precoder must be calculated every 1 to 2 milliseconds to adapt to fast changing wireless communication channels.

To the best of our knowledge, this work is the first attempt in searching for a practically feasible means of computing OLP for CFmMIMO. In this paper, we model OLP as a node-level prediction task on a graph. We represent the wireless communication channel between each AP and user equipment (UE) as a graph node and encode its channel coefficient as a node feature. We define two types of edges: an edge of type-UE connects two nodes (i.e., channels) that are interfering, while a type-AP edge indicates that they share the same transmitter.

Our model, named OLP-GNN, takes as input the aforementioned graph and outputs a precoding matrix which is trained to approximate the optimal linear precoder. It is based on the graph transformer architecture \cite{shi2020masked}. To satisfy the stringent runtime requirement, our GNN model has to be small: 6 hidden layers and about 22.4k trainable parameters. Given this, we design problem-specific data preprocessing and postprocessing methods to improve OLP-GNN's accuracy.
The preprocessing step consists in converting the complex-valued channel coefficients into 4 real-valued components which are then used as input of OLP-GNN. The GNN then predicts 6 features for each node that are combined to obtain a complex-valued precoding matrix in the postprocessing. These features have physical and mathematical interpretations, e.g., signal strength, interference power and power budget constraint. The postprocessing step also ensures that each AP's power budget constraint is satisfied.

We show via simulations that our solution can compute substantially optimal precoders within the time budget of 1 to 2 milliseconds for up to 96 APs and 18 UEs. We compare the spectral efficiency of OLP-GNN to two practical precoders, Maximum Ratio Transmission (MR) and Zero Forcing (ZF), highlighting the performance gain achieved by making OLP computable in real-time. We also show that a single trained model generalizes well to various system sizes and scenarios, including line-of-sight (LoS), non-line-of-sight (NLoS), urban and rural radio propagation environments.

\section{Related Work}

\textbf{CFmMIMO precoder designs:}
There are many papers on CFmMIMO precoder designs. Unlike OLP, all variants of MR (also known as conjugate beamforming)~\cite{zhao2020,luo2022downlink,salaun2022gnn} and ZF~\cite{elina2017,li2023heterogeneous,hao2024joint} are sub-optimal. In this paper, we will include MR and ZF as baselines. The scalable CFmMIMO framework is introduced in~\cite{bjornson2020scalable} which requires all processing tasks to have finite complexity as the number of UEs increases. In such a framework, the computations have to be distributed. This is not the case of our approach. Indeed, OLP-GNN is computed in a centralized manner with full knowledge of all channels. 
\cite{du2021cell} proposed a combination of MR and ZF, with most APs doing decentralized MR to minimize the front-haul burden.
\cite{hao2024joint} developed JointCFNet, a convolutional neural network for joint user association and power control with local partial protective ZF.
\cite{miretti2022} considered the uplink counterpart of OLP, i.e., the calculation of uplink joint optimal beamforming and power control for CFmMIMO.

\noindent
\textbf{Graphs neural networks for CFmMIMO:}
GNNs have been applied to the following optimization problems in CFmMIMO. Reference~\cite{salaun2022gnn} studied the downlink power control assuming MR precoding, while~\cite{shen2022graph} solved the uplink pilot power control. The authors of~\cite{li2023heterogeneous} tackled the joint downlink and uplink power control in a full-duplex system assuming ZF precoding. SINRnet is proposed in~\cite{raghunath2024energy} to maximize the downlink energy efficiency with MR precoding in an unsupervised manner. \cite{ranashinghe2021} optimizes the AP selection.
The graph structures in~\cite{salaun2022gnn,raghunath2024energy,eisen2020optimal} are similar to ours where nodes encode channels. However, their node features represent average channel amplitudes (real-valued), while we use instantaneous channel coefficients (complex-valued). The interference graph in~\cite{eisen2020optimal} can be seen as a special case of our graph where each AP serves exactly one UE, and only UE-type edges are considered. Other papers follow a different but common graph construction for wireless systems where each node is either an AP or a UE~\cite{ranashinghe2021,shen2022graph,li2023heterogeneous}.

\section{System Model and Notation} \label{sec:system_model}
\subsection{Notation}
 
We write vectors with bold font lowercase letters, e.g., $\bv$, and matrices with bold font capital letters, e.g., $\bA$. Superscripts $^{\sst{T}}$ and $^*$ denote respectively the transpose and complex conjugate transpose of a matrix. Thus, $^{\sst{T}*}$ and $^{*\sst{T}}$ denote the un-transposed conjugate. 
All vectors are assumed to be column vectors. 
For $\bv \in \mathbb{C}^K$, diag($\bv$) $\in \mathbb{C}^{K \times K}$ denotes the diagonal matrix  with $\bv$ as diagonal values. 
For $\bA \in \mathbb{C}^{K\times K}$, diag($\bA$) $\in \mathbb{C}^K$ denotes the diagonal matrix whose diagonal is the diagonal of matrix $\bA$.
$\|\cdot\|_2$ and $\|\cdot\|_{\infty}$ denote the $2$-norm and infinity norm respectively. 
For $\bA \in \mathbb{C}^{M\times K}$, $\Bar{\ba}_m$ denotes the $m$-th row of $\bA$.
For two matrices $\bA$ and $\bB$ of compatible sizes, [$\bA,\bB$] denotes their concatenation along the second dimension.
$\bI_K$ is the $K$-dimensional identity matrix. $\mathbb{R}^+$ denotes the set of all real positive numbers.
Let $x\in\mathbb{C}$ be a complex number, we denote its magnitude (absolute value) by $|x|$ and its phase by $\mbox{phase}(x)$. We define $[K]=\{1,\cdots,K\}$.

\subsection{Cell-Free Massive MIMO}
We consider a CFmMIMO system where $M$ APs transmit simultaneously to $K$ UEs in the downlink. A fundamental assumption of massive MIMO is that $M$ is greater than $K$~\cite{marzetta2016fundamentals}. The channel between any AP $m\in [M]$ and UE $k\in [K]$ is characterized by a complex channel coefficient $g_{m,k}\in\mathbb{C}$. The matrix of all channel coefficients is called \textit{channel matrix} and is denoted by $\bG \in \mathbb{C}^{M \times K}$. We have
 \begin{equation*}
    \bG=\begin{pmatrix}
g_{1,1} & \cdots & g_{1,K}\\
\vdots & \vdots & \vdots \\
g_{M,1} & \cdots & g_{M,K}
\end{pmatrix}= \begin{pmatrix}
\bg_{1} & \cdots & \bg_{K}
\end{pmatrix}= \begin{pmatrix}
\bar{\bg}_{1}^{\sst T}\\ \vdots \\ \bar{\bg}_{M}^{\sst T}
\end{pmatrix},
 \end{equation*}  
 where $\bg_{k} \in
 \mathbb{C}^{M}$ is the channel vector between the $k$-th UE and all $M$ APs, and $\bar{\bg}_{m} \in
 \mathbb{C}^{K}$ the channel vector between the $m$-th AP and all $K$ UEs.
The APs are connected to a central controller that collects the channel state information (CSI) which gives us $\bG$. The computations and neural network inferences presented in this paper are performed on this central controller, then the results are sent to each AP.

Let $\bx \in \mathbb{C}^{K}$ be the signal received by the $K$ users. It can be modelled as:
 \begin{equation}\label{eq:xMIMOequation}
     \bx=\bG^{\sst T}(\sqrt{\rho_{d}}\bs)+\bw,
 \end{equation}
 where $\rho_{d}$ is the downlink signal to noise ratio (SNR) for each AP, $\bs \in \mathbb{C}^{M}$ is the power normalized precoded signal to be transmitted by the $M$ APs and $\bw \in \mathbb{C}^{K}$ is a circularly-symmetric Gaussian noise vector.
The APs are subject to a power constraint set to
\begin{equation}\label{eq:spowerconstraint}
     \|\mathbb{E}(\bs^{*T} \odot \bs) \|_{\infty} \leq 1,   
\end{equation}
with $\mathbb{E}$ the expectation and $\odot$ the element wise multiplication. 
\subsection{Precoding Matrix and Downlink SINR Calculation} \label{section:SINR_calcul}

We denote by $\bq \in \mathbb{C}^K$ the users' message-bearing symbols to be transmitted. We assume, as in \cite{yang2021cell}, that $\bq$ has zero mean, unit variance and that the symbols are uncorrelated between users such that the following holds
\begin{equation} \label{eq:quncorrelated}
    \mathbb{E}(\bq \bq^{*})=\bI_{K}.
\end{equation} 
As highlighted in equation~\eqref{eq:xMIMOequation}, the signal to be transmitted by the APs is $\bs \in \mathbb{C}^M$. Therefore $\bq$ must be converted from the user data symbols space $\mathbb{C}^K $ to the precoded signals space $\mathbb{C}^M$.
This is done with a linear precoding matrix $\Delta$ as follows
\begin{equation} \label{eq:s_from_delta}
    \bs=\Delta \bq, \quad \text{where} \quad \Delta=\begin{pmatrix}
    \delta_{1,1} & \cdots & \delta_{1,K}\\
    \vdots & \vdots & \vdots \\
    \delta_{M,1} & \cdots & \delta_{M,K}
    \end{pmatrix}
    \in \mathbb{C}^{M \times K}.
\end{equation}

The assumption on $\bq$ in eq.~\eqref{eq:quncorrelated} combined with eq.~\eqref{eq:spowerconstraint} and eq.~\eqref{eq:s_from_delta} imposes the following power constraints on $\Delta$
\begin{equation}\label{eq:delta_power_constraint}
      \forall\, m\in [M], \quad\|\bar{\bdelta}_{m}\|_{2} \leq1,  
\end{equation}
where $\bar{\bdelta}_{m}=(\delta_{m,1},\cdots, \delta_{m,K})^{\sst T}\in \mathbb{C}^K$. 

From eq.~\eqref{eq:xMIMOequation} and eq.~\eqref{eq:s_from_delta} we get $\bx=\sqrt{\rho_d}\bG^{\sst T}\Delta\bq+\bw$,
which allows us to express the signal received at the $k$-th user as
\begin{equation*}
x_k
=\sqrt{\rho_d}\bg_k^{\sst T}\Delta\bq+w_k
=\sqrt{\rho_d}\bg_k^{\sst T}\bdelta_k q_k + \sqrt{\rho_d}\sum_{l\neq k}{\bg_k^{\sst T}\bdelta_l q_l}+w_k,
\end{equation*}
where $\bdelta_{k}=(\delta_{1,k},\cdots, \delta_{M,k})^{\sst T} \in \mathbb{C}^M$.
We know from eq.~\eqref{eq:quncorrelated} that a signal emitted for a specific user $k$ is uncorrelated with interfering signals intended for other users. Similarly, white additive noise is uncorrelated with both intended and interfering signals. Since the intended signal, interfering signals and noise are mutually uncorrelated, we can calculate their contribution to power separately.
Hence, the power of the signal $x_k$ received by the $k$-th user can be written as
$\mathbb{E}(x_k^*x_k)=\rho_d|\bg_k^{\sst T}\bdelta_k |^2+\rho_d\sum_{l\neq k}|\bg_k^{\sst T}\bdelta_l|^2+1$,
with the following terms:
\begin{itemize}
    \item Signal power (SP): $\rho_d|\bg_k^{\sst T}\bdelta_k |^2$ is the power of the signal intended for user $k$.
    \item Interference power (IP): $\rho_d\sum_{l\neq k}|\bg_k^{\sst T}\bdelta_l|^2 $.
    \item Noise power (NP) is equal to $1$.
\end{itemize}
The \textit{signal to interference plus noise ratio} (SINR) of a user $k$ is defined as the ratio between its intended signal power and the interference power plus noise power. Thus, the SINR of user $k$ can be calculated as
\begin{equation}
    \mbox{SINR}_k=\frac{\text{SP}}{\text{IP}+\text{NP}}=\frac{\rho_d|\bg_k^{\sst T}\bdelta_k |^2}{1+\rho_d\sum_{l\neq k}\\|\bg_k^{\sst T}\bdelta_l|^2}. \label{eq:sinr_delta}
\end{equation}
Equation~\eqref{eq:sinr_delta} can be expressed otherwise by introducing the following matrix $\bA\in\mathbb{C}^{K\times K}$ which combines both effects of precoding and channel propagation.
\begin{equation} \label{eq:def_A}
    \bA=\begin{pmatrix}
a_{1,1} & \cdots & a_{1,K}\\
\vdots & \vdots & \vdots \\
a_{K,1} & \cdots & a_{K,K}
\end{pmatrix}=\bG^{\sst T}\Delta.
\end{equation}
Equation~\eqref{eq:sinr_delta} then becomes
\begin{equation}
     \mbox{SINR}_{k}=\frac{\rho_{d}|a_{k,k}|^{2}}{1+\rho_{d}\sum_{l\neq k}|a_{k,l}|^{2}}.\label{eq:sinr_a}
\end{equation}

\subsection{Optimal Linear Precoding} \label{section:OLP}
We define the optimal linear precoding, denoted by $\Delta_{\sst{OLP}}$, as the solution to the following max-min SINR problem.
\begin{equation*}\tag{$\mathcal{P}$}\label{P}
\begin{aligned}
& \underset{\Delta}{\mbox{max}}\quad\underset{k}{\mbox{min}}
& & \mbox{SINR}_k, \\
& \text{subject to}
& & \|\bar{\bdelta}_{m}\|_{2} \leq1, \quad \forall m.
\end{aligned}
\end{equation*}
The objective of this problem is to maximize the minimum $\mbox{SINR}$ among all UEs while satisfying the power constraint~\eqref{eq:delta_power_constraint}. Note that the objective is a function of $\Delta$ as shown in~\eqref{eq:s_from_delta} and~\eqref{eq:sinr_delta}. We will explain in the following paragraphs how the optimal $\Delta_{\sst{OLP}}$ can be obtained by a combination of bisection search and second order cone programming (SOCP) feasibility search.

We first focus on the following feasibility subproblem. Given a threshold value $t_{\mbox{\scriptsize{SINR}}}$, it consists in checking whether there exists a feasible solution such that
$\min_k\,\mbox{SINR}_{k}\geq t_{\sst SINR}$.
This inequality can be expanded with eq.~\eqref{eq:sinr_a} as
\begin{equation*}
\forall k \in [K],\quad \frac{\rho_{d}|a_{k,k}|^{2}}{1+\rho_{d}\sum_{l\neq k}|a_{k,l}|^{2}} \geq t_{\sst{SINR}},
\end{equation*}
which is equivalent to
\begin{equation}
|a_{k,k}|^{2}\geq t_{\sst{SINR}} \left(\frac{1}{\rho_d}+\sum_{l\neq k}|a_{k,l}|^{2}\right).\label{eq:sinr_treshold}
\end{equation}
By introducing the following matrix
$\Tilde{\bA}=\begin{bmatrix}
\bA-\mbox{diag}(a_{1,1}, \cdots, a_{K,K}); \frac{1}{\sqrt{\rho_d}}\bOne_{K \times 1}
\end{bmatrix}$ in $\mathbb{C}^{K\times (K+1)}$,
inequality~\eqref{eq:sinr_treshold} can be simplified as
\begin{equation}
\forall k \in [K],\quad |a_{k,k}| \geq \sqrt{t_{\sst{SINR}}}\|\Tilde{\bf{a}}_k\|_2, \label{eq:sinr_treshold_tilde_a}
\end{equation}
where $\Tilde{\bf{a}}_k=(\Tilde{a}_{k,1}, \cdots, \Tilde{a}_{k,K+1})\in \mathbb{C}^{K+1}$.
We note that the left term in equation~\eqref{eq:sinr_treshold_tilde_a} is convex due to the absolute value $|\cdot|$. It has to be concave to match a standard form of SOCP constraint. To this end, we will restrict the set of possible values for $a_{k,k}$.

We derive from eq.~\eqref{eq:def_A} that $\Delta$ can be written as a function of $\bA\in \mathbb{C}^{K\times K}$ and an arbitrary matrix $\bU \in \mathbb{C}^{M \times K}$ such that
\begin{equation}\label{eq:delta_olp}
    \Delta=\bG^{\dag}\bA+P_{\bG^{\sst{T}}}\bU,
\end{equation}
where $\bG^{\dag}=\bG^{\sst T*}\left(\bG^{\sst T}\bG^{\sst T*}\right)^{-1}$ is the Moore-Penrose pseudo-inverse of $\bG^{\sst T}$, and $P_{\bG^{\sst T}}=\bI_{M}-\bG^{\dag}\bG^{\sst T}$ is the orthogonal projection onto the null space of $\bG^{\sst T}$.

We can see that the right multiplication of $\Delta$ by a matrix $\mbox{diag}(e^{i\theta_1}, \cdots , e^{i\theta_K})$ does not change its 2-norm, thus leaving the right hand side of~\eqref{eq:sinr_treshold_tilde_a} unchanged. Besides, if $(\bA, \bU)$ satisfies inequality~\eqref{eq:sinr_treshold_tilde_a} and the power constraint~\eqref{eq:delta_power_constraint}, then so does $\left(\bA\mbox{diag}(e^{i\theta_1}, \cdots , e^{i\theta_K}),\bU\mbox{diag}(e^{i\theta_1}, \cdots , e^{i\theta_K})\right)$. Therefore, if we multiply $\bA$ and $\bU$ by $\mbox{diag}(e^{i\theta_1}, \cdots , e^{i\theta_K})= \mbox{diag}(-\mbox{phase}(\bA)) $, we can restrict our search to positive real values of $a_{k,k}$, for all $k$, instead of complex values. With this assumption, the max-min problem~\eqref{P} can be reformulated as:
\begin{equation*}\tag{$\mathcal{P'}$}\label{P'}
\begin{aligned}
& \underset{\Delta}{\mbox{max}} && t_{\sst{SINR}}, \\
& \text{subject to} && a_{k,k} \geq \sqrt{t_{\sst{SINR}}}\|\Tilde{\bf a}_k\|_2, &&&  \forall k, \\
& && a_{k,k}\in\mathbb{R}^{+}, &&&  \forall k,\\
& && \| \bar{\bdelta}_m \|_2 \leq 1, &&&  \forall m.
\end{aligned}
\end{equation*}
The constraints of~\ref{P'} are written in a standard form suitable for SOCP. Hence, for any value of $t_{\sst{SINR}}$, SOCP can be applied to check the feasibility of the constraints. A bisection search can be used on top of SOCP to find the maximum value of $t_{\sst{SINR}}$. We will refer to this method as B-SOCP throughout the paper.

\subsection{Zero Forcing and Maximum Ratio Precoding}
Other linear precoding schemes are often considered, which are by definition sub-optimal compared to OLP but less costly to solve. In this paper, we will compare our solution to two precoding schemes, namely Zero Forcing (ZF) and Maximum Ratio Transmission (MR). ZF minimizes the interference while MR maximizes the signal. The optimal is a trade-off between these two extremes. The ZF precoder has a closed-form expression which can be computed by matrix multiplications and inversions~\cite{elina2017}. For MR, the max-min objective can be achieved by solving a B-SOCP problem similar to~\ref{P'}~\cite{ngo2017}.

It is known that for some regimes (e.g., high signal-to-noise, large number of APs), ZF outperforms MR whereas for other regimes it is the opposite.
This can be seen in section~\ref{sec:numerical_results} for example, where MR achieves higher spectral efficiency than ZF in Fig.~\ref{fig:se_nlos_rural_32_16} and ZF beats MR in Fig.~\ref{fig:se_nlos_rural_96_36}. In all scenarios, OLP significantly outperforms both MR and ZF. However, the computational complexity of B-SOCP makes it unsuitable for real world systems with millisecond-scale runtime requirements. This shows the importance of developing an approximation of OLP with several order of magnitudes faster runtimes.

\section{Graph Neural Network}
In this section, we describe our solution, named OLP-GNN, to tackle the max-min SINR problem~\ref{P}. OLP-GNN is trained with OLP data obtained by running B-SOCP in a simulated environment. The objective is to approximate the performance of OLP with a low and practical computational complexity.

\subsection{Graph Representation}\label{section:graph_rep}
The input and output of our max-min problem~\ref{P} are respectively the channel matrix $\bG$ and the precoding matrix $\Delta_{\sst{OLP}}$. 
One can see that for any permutation applied to the rows and/or columns of $\bG$, the same permutation is applied to the optimal solution $\Delta_{\sst{OLP}}$. Thus the problem is independent from the row and column indexing. This property is called \textit{permutation equivariance} and GNNs are known to satisfy this property, which make them suitable for our problem.

To train a GNN the input channel matrix $\bG$ and output precoding matrix $\Delta_{\sst OLP}$ must be represented as graphs. We define a directed graph as $(V,E)$ where $V$ is the set of nodes and $E$ the set of directed edges. We define a node as a pair $(m,k)\in [M]\times[K]$, thus $V$ has $M\times K$ nodes. We also define $\pi$, a bijection from the set of pairs $(m,k)$ to the set of nodes $V=[MK]$, that associates to each $(m,k)$ pair a node index $i \in V$, such that $\pi(m,k)=i$.

In our problem when two UEs share the same AP, i.e., they are in the same row of $\Delta_{\sst{OLP}}$, they mutually influence each other through the power constraint of equation~\eqref{eq:delta_power_constraint}. Similarly, when two APs serve the same UE $k$, i.e., they are on the same column of $\Delta_{\sst{OLP}}$, they both have an impact on the calculation of $\mbox{SINR}_k$ in equation~\eqref{eq:sinr_delta}. To encode these properties in our graph, we consider two types of edges.
We set an edge between $i,j \in V$ if and only if they share a common AP or a common UE. We denote this edge by $e=(i,j)$ and its type by $\mbox{type}(e) \in \{\text{AP}, \text{UE}\}$ depending on whether $i$ and $j$ share a common AP or UE. The graph does not have self loop, i.e., $ \forall i \in V, (i,i) \notin E$. This formally translates to for all $m, m'\in [M]$, $m\neq m'$, and for all $k, k'\in [K]$, $k \neq k'$, we have:
\begin{enumerate}[label=\roman*.]
    \item $e=(\pi(m,k),\pi(m,k')) \in E$ and $\mbox{type}(e)=\text{AP}$,
    \item $e=(\pi(m,k),\pi(m',k)) \in E$ and $\mbox{type}(e)=\text{UE}$.
\end{enumerate}

The heterogeneous nature of $E$ allows a GNN to process differently the information on a node and its neighbors based on their edge type. 
Thus, for each node $i\in V$, we define two disjoint sets of neighbors depending on their edge type:
$\mathcal{N}_{\sst{AP}}(i)= \{j \in V \, | \, (i,j) \in E \text{ and } \mbox{type}\!\left((i,j)\right)=\text{AP}\}$ and $\mathcal{N}_{\sst{UE}}(i)= \{j \in V \, | \, (i,j) \in E \text{ and } \mbox{type}\!\left((i,j)\right)=\text{UE}\}$.

\begin{wrapfigure}[10]{r}{0.5\textwidth}
    \centering
    \vspace{-0.66cm}
    \resizebox{0.45\textwidth}{!}{%
    \definecolor{bleu}{rgb}{0.0, 0.0, 1.0}
\definecolor{vert}{rgb}{0.0, 0.5, 0.0}
\newcommand{\neighborsUE}[1]{\mathcal{N}_{\text{UE}}(#1)}
\newcommand{\neighborsAP}[1]{\mathcal{N}_{\text{AP}}(#1)}
\begin{tikzpicture}
    \def \offset {360/5.4}
    \def \radius {4.5cm}
    \def \nodesize {1.6cm}
    \def \numnodes {8}
    \tikzstyle{self_node}=[circle, fill=white, line width=0.5mm,draw,minimum size=\nodesize]
	\tikzstyle{ue_node}=[circle,line width=0.5mm,draw=vert,fill=vert!20,minimum size=\nodesize]
	\tikzstyle{ap_node}=[circle,line width=0.5mm,draw=bleu,fill=bleu!20,minimum size=\nodesize]
	\tikzset{edge/.style = {->}}

    \begin{scope}

    \draw [line width=1mm,draw=vert!80, fill=vert!03,dashed](180:1.22*\radius)--(0,0)--(91:1.22*\radius) arc (91:180:1.22*\radius);
    
    \draw [line width=1mm,draw=bleu!80, 
    fill=bleu!03,dashed](360:1.22*\radius)--(0,0)--(449:1.22*\radius) arc (449:360:1.22*\radius);
    \node[self_node] (center) at (0,0) {$\pi(m,k)$};
    \end{scope}

    \begin{scope}
    \node[ue_node] (v1) at (\offset + 360/\numnodes * 1:\radius) {$\pi(1,k)$};

    \node[ue_node,label={[label distance=2.6cm]above:\Large\textcolor{vert}{$\neighborsUE{i}$}}] (v2) at (\offset + 360/\numnodes * 2.2:\radius) {$\pi(M,k)$};
    
    \path (v1) -- node[auto=false, rotate=45]{\Large$\bullet\bullet\bullet$} (v2);
    
    \end{scope}

    \begin{scope}
    \node[ap_node,label={[label distance=2.6cm]above:\Large\textcolor{bleu}{$\neighborsAP{i}$}}] (v3) at (\offset + 360/\numnodes * 6.8:\radius) {$\pi(m,K)$};

    \node[ap_node] (v4) at (\offset + 360/\numnodes * 8:\radius) {$\pi(m,1)$};

    \path (v3) -- node[auto=false, rotate=135]{\Large$\bullet\bullet\bullet$} (v4);

    \end{scope}
    
    \draw[edge,draw=vert!80,line width=0.75mm] (center) to node[sloped,xshift=1mm,yshift=-3mm]{\textcolor{vert}{$\mbox{type} =$ UE}}(v1);
    
    \draw[edge,draw=vert!80,line width=0.75mm] (center) to node[sloped,xshift=1mm,yshift=3mm]{\textcolor{vert}{$\mbox{type} =$ UE}}(v2);

    \draw[edge,draw=bleu!80,line width=0.75mm] (center) to node[sloped,xshift=1mm,yshift=3mm]{\textcolor{bleu}{$\mbox{type} =$ AP}}(v3);

    \draw[edge,draw=bleu!80,line width=0.75mm] (center) to node[sloped,xshift=-1mm,yshift=-3mm]{\textcolor{bleu}{$\mbox{type} =$ AP}}(v4);
      
\end{tikzpicture}
    }%
  \vspace{-0.24cm}
  \caption{Neighbors and outgoing edges of a typical node $\pi(m,k) = i \in V$}\label{fig:typical_node}
\end{wrapfigure}
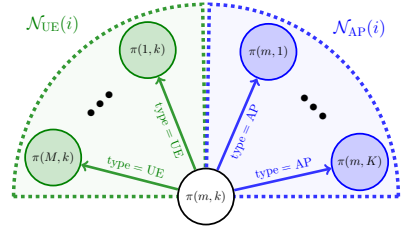

Figure~\ref{fig:typical_node} illustrates the neighbors of a typical node. We note that for each node $i \in V$, $\mathcal{N}_{\sst{AP}}(i)$ has $K-1$ elements and $\mathcal{N}_{\sst{UE}}(i)$ has $M-1$ elements, for a total of $M+K-2$ neighbors. In other words, each node  has $M+K-2$ outgoing edges, and the same number of incoming edges. Since the graph contains $MK$ nodes, the total number of edges is $MK(M+K-2)$.

\subsection{Data Preprocessing and Postprocessing}
In this subsection, we consider a typical node $i \in V$ corresponding to the channel between AP $m$ and UE $k$, i.e., $\pi(m,k) = i$. At each iteration $t$ of the GNN, node $i$ is associated with a tensor $h_i(t)$, called \textit{node feature}. $h_i(0)$ is its input feature and $h_i(T)$ is its output feature, where $T$ is the number of iterations/layers. We assume that all tensors in OLP-GNN are real-valued. Since the input and target data are complex matrices, we decompose them into their magnitude and phase components. Another possibility is to decompose them into their real and imaginary parts. However, we found magnitude-phase representation to be more fitting, probably due to its much smaller ranges.

We see from eq.~\eqref{eq:delta_olp} that the precoding matrix depends directly on $\bG^{\dag}$. During the development of OLP-GNN, we observed that learning the pseudo-inverse $\bG^{\dag}$ gave an unsatisfactory accuracy and a poor generalization to different number of APs and UEs.
Indeed, the performance drops by $5-20\%$ on the validation datasets without $\bG^{\dag}$ in the input of our $T=6$ layers model.
Thus, we decided to compute $\bG^{\dag}$ beforehand using fast numerical methods and include it in OLP-GNN's input. Therefore, $h_i(0)$ contains 4 elements which are the magnitudes and phases of $\bg_{m,k}$ and $\bg^{\dag}_{m,k} = (\bG^{\dag})_{m,k}$, i.e., $h_i(0)=\left(|\bg_{m,k}|,\,\mbox{phase}(\bg_{m,k}),\,|\bg^{\dag}_{m,k}|,\,\mbox{phase}(\bg^{\dag}_{m,k})\right)$.

We split the target precoding matrix $\Delta_{\sst{OLP}}$ in three components $\bG^{\dag}\mbox{diag}(\bA)$, $\bG^{\dag}(\bA-\mbox{diag}(\bA))$, $P_{\bG^T}\bU$. Therefore the output and target feature tensor contains the following $6$ terms:
magnitude and phase of $(\bG^{\dag}\mbox{diag}(\bA))_{m,k}$, 
magnitude and phase of $(\bG^{\dag}(\bA-\mbox{diag}(\bA))_{m,k}$, 
magnitude and phase of $(P_{\bG^T}\bU)_{m,k}$.
This split is motivated by the distinct physical meanings of these terms. The diagonal elements $(a_{k,k})_{k \in [K]}$ of $\bA$ represent user $k$'s useful signal. The non-diagonal elements $(a_{l,k})_{k \neq l}$ correspond to the interfering signals intended for user $k$ but received by another user $l$. In this sense, $\bA$ fully characterizes the SINR. On the other hand, $P_{\bG^{\sst{T}}}\bU$ only influences the power constraint without changing the SINRs. 

For OLP-GNN to extract useful information from the features, they must be of the same order of magnitude. The absolute values of the input, output and target features range over several orders of magnitude. Therefore, we apply a $\log_2$ transformation to all absolute values (magnitude terms).
As an example, we typically have $ 10^{-15} \leq |g_{m,k}| \leq 10^{-5}$, hence $\log_2(|g_{m,k}|)$ belongs to $[-50, -16]$. We do not apply a $\log_2$ transformation to the phase terms as they are already in a small range between $0$ and $2\pi$. All the features are then normalized to have zero mean and unitary variance.

Let $\by_1$, $\by_2$, $\by_3$ be the OLP-GNN predictions of the aforementioned three terms: $\bG^{\dag}\mbox{diag}(\bA)$, $\bG^{\dag}(\bA-\mbox{diag}(\bA))$, $P_{\bG^{\sst{T}}}\bU$. They are obtained by de-processing the output tensors $h_i(T)$ for all nodes $i \in V$. We apply the following postprocessing to impose some desired properties on the output
\begin{equation*}
    \begin{cases}
        \by'_{1}= \bG^{\dag}\mbox{real}(\mbox{diag}(\bG^{\sst{T}} \by_1)), & \\
        \by'_{2}= \bG^{\dag}(\bG^{\sst{T}}\by_2-\mbox{diag}(\bG^{\sst{T}}\by_2)),&\\
        \Delta = \by'_{1}+\by'_{2}+\by_3.
    \end{cases}
\end{equation*}
As we have $\bI_{K}=\bG^{\sst T}\bG^{\dag}$, the postprocessing on $\by'_{1}$ ensures that $\bG^{\sst T}\by'_1$ is a real diagonal matrix. The postprocessing on $\by'_{2}$ enforces that the diagonal elements of $\bG^{\sst T}\by'_2$ are equal to zero.
Once this is applied on the output features, we further impose the power constraint~\eqref{eq:delta_power_constraint} by applying the following projection
\begin{equation*}
    \forall m\in [M],
    \text{ if } \|\bar{\bdelta}_m\|_2 \geq 1\text{ then }\bar{\bdelta}_m\leftarrow \frac{\bar{\bdelta}_m}{\|\bar{\bdelta}_m\|_2}.
\end{equation*}
$\Delta$ obtained by the sum of components $\by'_{1}$, $\by'_{2}$, $\by_{3}$, and after the above projection is applied is the predicted precoding matrix of OLP-GNN. 

\subsection{Structure of the Neural Network} \label{section:structure_gnn}
Let $\mathcal{L}$ be the linear operator. For the sake of clarity, the linear layers in this section will be written with different subscripts and superscripts, e.g., $\mathcal{L}^1_{\sst{AP},t}$, $\mathcal{L}^4_{\sst{UE},t}$. This is done to indicate that they are applied on different edge types ($\mbox{AP}$ or $\mbox{UE}$), at different iterations $t \in \{0, \dots, T\}$ and they do not share any trainable parameter.

\begin{figure*}[t]
    \centering
\includegraphics[width=1.0\textwidth]{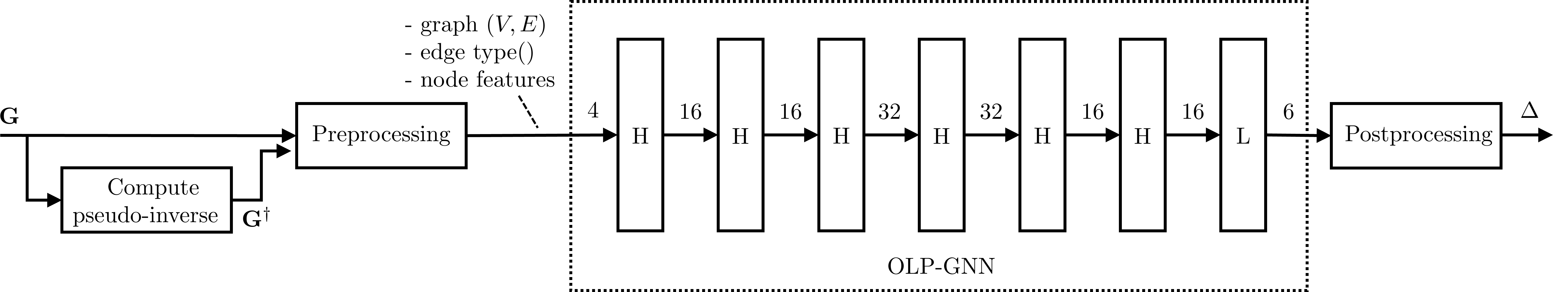}
    \caption{Structure of OLP-GNN. `H' refers to the hidden attention layer, and `L' is the final linear layer. The number between each layer represents the node feature size.}
    \label{fig:image_gnn}
\end{figure*}

For each node $i \in V$, its feature $h_i$ is updated based on itself and its direct neighbors at the previous step. Thus, $h_i(t+1)$ is a function of $h_i(t)$ and $h_j(t)$, $\forall j\in \mathcal{N}_{\sst{AP}}(i) \cup \mathcal{N}_{\sst{UE}}(i)$, updated according to the following rule
\begin{equation} \label{eq:update_rule}
    h_i(t+1)=\mbox{Norm}\left(\mbox{ReLU}(f_{\sst{AP},t}(i)+f_{\sst{UE},t}(i))\right),
\end{equation}
where Norm denotes the layer normalization and ReLU the rectified linear unit activation function.
For $i,j \in V$, $\bullet \in \{\mbox{AP},\mbox{UE}\}$ and $t \in \{0, \dots, T-1\}$, function $f_{\bullet,t}$ implements the graph transformer of~\cite{shi2020masked} with a single attention head. It is defined as 
\begin{equation}\label{eq:f_def}
    f_{\bullet,t}(i) =\mathcal{L}^1_{\bullet,t}(h_i(t))+ \sum_{j \in \mathcal{N}_{\bullet}(i)}\alpha_{\bullet,t}(i,j) \times \mathcal{L}^2_{\bullet,t}(h_j(t)).
\end{equation}  
The attention coefficient $\alpha_{\bullet,t}(i,j)$ is equal to $\langle \mathcal{L}^3_{\bullet,t}(h_i(t)),\mathcal{L}^4_{\bullet,t}(h_j(t))\rangle$ divided by $\sum_{u \in \mathcal{N}_{\bullet}(i)}\langle\mathcal{L}^3_{\bullet,t}(h_i(t)),\mathcal{L}^4_{\bullet,t}(h_u(t))\rangle$,
where $\langle x,y \rangle =\exp\left(\frac{x^{\sst{T}}y}{\sqrt{d}}\right)$ is the exponential scalar product~\cite{vaswani2017attention} and $d$ the size of tensors $x$ and $y$.

In this context, the attention is an efficient mechanism to select which neighbors have the most impact on improving the node's OLP prediction task according to their channels. It is important to note that a permutation of the nodes indices does not change the output value due to the summation used as an aggregator in equation~\eqref{eq:f_def}. As a consequence, the update rule~\eqref{eq:update_rule} satisfies the permutation equivariance property of our problem.

Figure~\ref{fig:image_gnn} shows the structure of our solution. OLP-GNN has $T=6$ hidden attention layers implementing the update rule~\eqref{eq:update_rule}. We choose this value since increasing the number of layers to $7$ does not improve the average performance by more than $1\%$ while increasing the runtime by $5-15\%$. The final iteration is a simple linear layer. OLP-GNN has a total of 22.4k trainable parameters.

\subsection{Training and Loss Function}\label{sec:training}
For training OLP-GNN, we generate data from two environments: free space 60GHz LoS and urban 2GHz NLoS. Each environment is simulated with the following number of APs and UEs placed randomly in a circular area of 500m radius: $(M,K)=(32, 6)$, $(32, 9)$, $(64, 9)$, $(64, 18)$. For each channel matrix $\bG$, we compute the corresponding target $\Delta_{\sst{OLP}}$ using B-SOCP. Each of the above 8 datasets has 10k samples, for which 9k are used for training, 500 for validation and 500 for testing. In summary, there is a total of 72k training samples. Details about the simulation settings can be found in the next section. Moreover, additional test datasets will be introduced there to evaluate the generalizability of OLP-GNN.

We consider the mean square error loss of the per-user SINR to train our model, i.e., 
$\sum_{k=1}^{K}(\mbox{SINR}^*_k-\mbox{SINR}_k)^2/K$, where $\mbox{SINR}^*_k$ is the target SINR value and $\mbox{SINR}_k$ is the SINR value predicted by OLP-GNN for user $k$. The SINRs used in the loss are expressed in dB.
We use the Adam optimizer \cite{kingma2014adam} for the training with a learning rate of $7\times 10^{-4}$, a batch size of 16, and 1000 epochs.

\section{Numerical Results}~\label{sec:numerical_results}
In this section, we describe in detail the simulation used to generate training, validation and test data for OLP-GNN. We then analyze our solution in terms of spectral efficiency, computational complexity and runtime. We also compare OLP-GNN to the classical MR and ZF precodings, as well as the target OLP.

\subsection{Simulation Parameters and Performance Metrics}
We simulate three CFmMIMO environments, namely 60GHz LoS, urban 2GHz NLoS and rural 450MHz NLoS. The LoS model is identical to the one in~\cite{yang2021cell} for 60GHz carrier frequency. The NLoS environments are modeled following the ITU-R~\cite{ITU-R2009} recommendations. Specifically, we consider the urban macro and rural macro NLoS radio propagation models with respectively 2GHz and 450MHz carrier frequencies. We deliberately choose different carrier frequencies to show the generalizability of our solution. We consider a bandwidth of 20 MHz for all three environments.

For each environment we define 24 scenarios with different number of APs ($M=24\ldots96$) and UEs ($K=4\ldots36$) as summarized in Table~\ref{tab:gnn_vs_socp_time}. The APs and UEs are randomly positioned inside a circular area of radius 500m (4km for the rural environment). As explained in section~\ref{sec:training}, four LoS datasets and four urban NLoS datasets are used for training with 9k training samples each. The rest of the LoS and urban NLoS scenarios each has 500 validation samples for hyperparameter tuning and 500 test samples. Finally, the rural NLoS scenarios are dedicated exclusively for testing with 500 samples each.

In contrast to~\cite{elina2017,zhao2020,salaun2022gnn} which only consider large-scale fading to optimize the power control, precoding is done at a much shorter time scale which requires to account for fast fading.  Indeed, the channel coefficient $g_{m,k}$ is equal to the large-scale fading between the $m$-th AP and $k$-th UE multiplied by a fast fading term $\zeta_{m,k}$. Here, we assume the fast fading to be i.i.d. Rayleigh distributed, i.e., $\zeta_{m,k}=\left( x_1 + x_2 \mathrm{i}\right)/\sqrt{2}$,
where $x_1$ and $x_2$ are independent standard normal random variables and $\mathrm{i}$ denotes the imaginary unit. The magnitude of the complex random variable $\zeta_{m,k}$ follows Rayleigh distribution.

We define the \textit{spectral efficiency} (SE) in bit/s/Hz of user $k$ as $\mbox{SE}=\mbox{log}_2(1+\mbox{SINR}_k)$.
We introduce two metrics based on SE to study the performance of OLP-GNN, the \textit{performance loss at median} and the \textit{95\%-likely SE loss}. The performance loss at median refers to the relative difference in spectral efficiency between our solution and OLP taken at the median of their cumulative distribution functions (CDF). The 95\%-likely performance metric is the relative loss at the $5$-th percentile, thus indicating the coverage quality for 95\% of users.

The performances of our solution in terms of spectral efficiency, runtime and complexity will be compared to the optimal B-SOCP solution which defines the upper bound for spectral efficiency but achieved in a time consuming manner.

The simulations and algorithms are implemented in Python 3 and can be found at \url{https://github.com/Nokia-Bell-Labs/olp-gnn}. The optimal linear precoding matrix is obtained by solving problem~\ref{P'} using 
the MOSEK solver~\cite{andersen2003implementing} for SOCP combined with a bisection search as explained in section~\ref{section:OLP}. The bisection search terminates when precision $\epsilon=0.01$ is reached for all SINRs. The OLP-GNN is implemented in PyTorch 2 and is compiled with PyTorch's default backend TorchInductor for runtime measurements.

\newcommand{\columnlength}{1.5cm}
\begin{table*}[t]
    \centering
    \caption{OLP-GNN performance and runtimes. 
    }
    \label{tab:gnn_vs_socp_time}
    \resizebox{\columnwidth}{!}{
    \begin{tabular}{ ?c|c?>{\centering\arraybackslash}p{\columnlength}|>{\centering\arraybackslash}p{\columnlength}|>{\centering\arraybackslash}p{\columnlength}|>{\centering\arraybackslash}p{\columnlength}|>{\centering\arraybackslash}p{\columnlength}|>{\centering\arraybackslash}p{\columnlength}?c|c? }
        \thickhline
        \multicolumn{2}{?c?}{Graph size} & \multicolumn{6}{c?}{Spectral efficiency loss compared to the optimal (\%)} & \multicolumn{2}{c?}{Runtime (ms)}
        \\
        \cline{1-10} \hphantom{.} \multirow{2}{*}{$M$} \hphantom{.} & \hphantom{.} \multirow{2}{*}{$K$} \hphantom{.}
        & \multicolumn{2}{c|}{LoS} & \multicolumn{2}{c|}{Urban NLoS} & \multicolumn{2}{c?}{Rural NLoS} & \multirow{2}{*}{average} & std
        \\
        \cline{3-8} & & median & 95\%-likely & median & 95\%-likely & median & 95\%-likely & & ($\times 10^{-3}$)
        \\ \thickhline
        
        $24$ & $4$ & $0.92$ & $1.26$ & $0.83$ & $1.87$ & $0.79$ & $2.99$ & \hphantom{.} $0.94$ \hphantom{.} & \hphantom{.} $7.25$ \hphantom{.}
        \\
		\hline
        $24$ & $5$ &  $0.74$ & $1.07$ & $0.72$ & $1.46$ & $1.22$ & $3.40$ &  $0.94$ & $7.10$
        \\
        \hline
        $24$ & $6$ &  $0.71$ & $1.34$ & $0.74$ & $3.49$ & $1.31$ & $2.77$ & $0.95$ & $7.75$
        \\
        \hline
        $24$ & $9$ & $0.86$ & $1.96$  & $1.20$ & $5.59$ & $2.67$ & $7.76$ & $0.95$ & $8.18$
        \\
        \hline
        $32$ & $4$ & $1.33$ & $1.65$ & $0.32$ & $0.16$ & $0.29$ & $-0.05$ & $0.95$  & $7.82$
        \\
        \hline
        $32$ & $6$ &  $0.50$ & $0.69$ & $0.27$ & $0.72$ & $0.76$ & $0.75$ & $0.96$ & $7.85$
        \\
        \hline 
        $32$ & $8$ &  $0.50$ & $0.80$ & $0.44$ & $0.62$ & $0.74$ & $1.10$ & $0.96$ & $7.86$
        \\
        \hline
        $32$ & $9$ & $0.64$ & $0.65$ & $0.55$ & $1.21$ & $0.06$ & $2.69$ & $0.96$ & $7.84$
        \\
        \hline
        $32$ & $12$ & $0.55$ & $0.88$ & $0.65$ & $1.18$ & $1.18$ & $4.99$ & $0.96$ & $7.55$
        \\
        \hline
        $32$ & $16$ & $1.17$ & $3.37$ & $0.35$ & $4.06$ & $2.21$ & $11.57$ & $0.97$ & $8.29$
        \\
        \hline
        $48$ & $8$ & $0.62$ & $0.76$ & $0.24$ & $0.64$ & $0.09$ & $0.33$ & $0.99$ & $7.59$
        \\
        \hline
        $48$ & $12$ & $0.56$ & $0.69$ & $0.56$ & $0.72$ & $0.47$ & $1.07$ & $1.02$ & $8.70$
        \\
        \hline
        $48$ & $16$ & $0.61$ & $0.69$ & $0.48$ & $0.09$ & $0.12$ & $1.43$ & $1.06$ & $7.65$
        \\
        \hline
        $48$ & $24$ & $1.31$ & $2.94$ & $0.07$ & $2.25$ & $0.40$ & $2.56$ & $1.24$ & $8.08$
        \\
        \hline
        $64$ & $6$ &  $0.94$ & $1.28$ & $0.59$ & $0.64$ & $0.46$ & $0.46$ & $0.99$ & $7.87$ 
        \\
        \hline
        $64$ & $9$ & $0.47$ & $0.83$ & $0.29$ & $0.90$ & $0.44$ & $0.04$ & $1.04$ & $6.99$
        \\
        \hline
        $64$ & $12$ & $0.56$ & $0.78$ & $0.51$ & $0.71$ & $0.29$ & $0.15$ & $1.09$ & $7.73$
        \\
        \hline
        $64$ & $18$ &  $0.58$ & $0.69$ & $0.54$ & $0.50$ & $0.24$ & $0.32$ & $1.28$ & $8.09$
        \\
        \hline
        $64$ & $24$ &  $0.44$ & $0.59$ &  $0.37$ & $0.63$ & $-0.18$ & $-0.92$ & $1.41$ & $7.96$
        \\
        \hline
        $64$ & $32$ & $1.70$ & $3.50$ & $-0.06$ & $1.73$ & $-0.49$ & $0.82$ & $1.55$ &  $8.54$
        \\
        \hline
        $96$ & $9$ & $0.79$ & $1.19$ & $0.46$ & $0.42$ & $0.37$ & $0.47$ & $1.19$ &  $7.45$
        \\
        \hline
        $96$ & $18$ & $0.69$ & $0.68$ & $0.48$ & $0.55$ & $0.41$ & $0.66$ & $1.59$ &  $8.33$
        \\
        \hline
        $96$ & $27$ & $0.72$ & $0.92$ & $0.57$ & $1.04$ & $0.26$ & $-0.04$ & $2.02$ &  $8.64$
        \\
        \hline
        $96$ & $36$ & $1.30$ & $2.52$ & $0.41$ & $0.95$  & $0.05$ & $-0.13$ & $2.53$ &  $9.67$
        \\
        \thickhline
    \end{tabular}
    }
\end{table*}

\subsection{Spectral Efficiency}

To highlight the generalizability of OLP-GNN, we test it on a rural NLoS environment. This is in contrast to the LoS and urban NLoS environments seen during training. The rural channel distribution differs from the training datasets due to the different radio propagation models, cell sizes and carrier frequencies employed.
The results are summarized in Table~\ref{tab:gnn_vs_socp_time}. Note that, the median or 95\%-likely SE loss written in the table can be negative since the precoder produced by OLP-GNN may give higher SE to some users at the expense of other users. In this case, the max-min objective is not reached and OLP-GNN necessarily under-performs at other parts of the CDF. For example, in the rural NLoS scenario with 64 APs and 24 UEs, OLP-GNN achieves negative median and 5-th percentile losses. 
Nevertheless, it is $0.20\%$ away from optimal at the 1st percentile of the CDF (not shown on the table).

\begin{figure}
\centering
\subfloat[LoS: 96 APs, 36 UEs\label{fig:se_los_96_36}]{
  \centering \includegraphics[width=0.48\textwidth, trim={0pt 10pt 10pt 10pt}, clip]{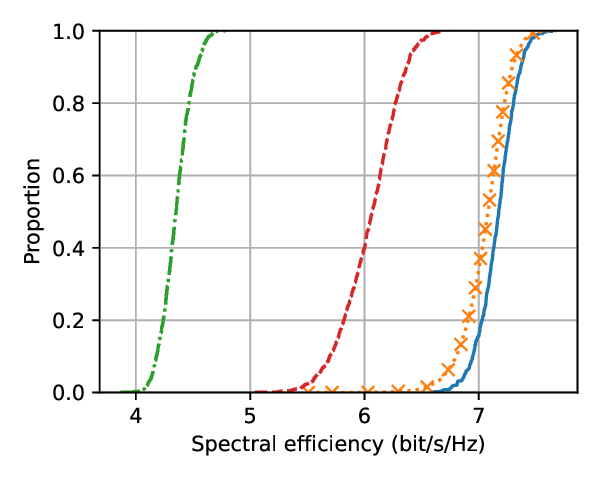}
}
\subfloat[Urban NLoS: 96 APs, 36 UEs\label{fig:se_nlos_96_36}]{
    \centering \includegraphics[width=0.48\textwidth, trim={0pt 10pt 10pt 10pt}, clip]{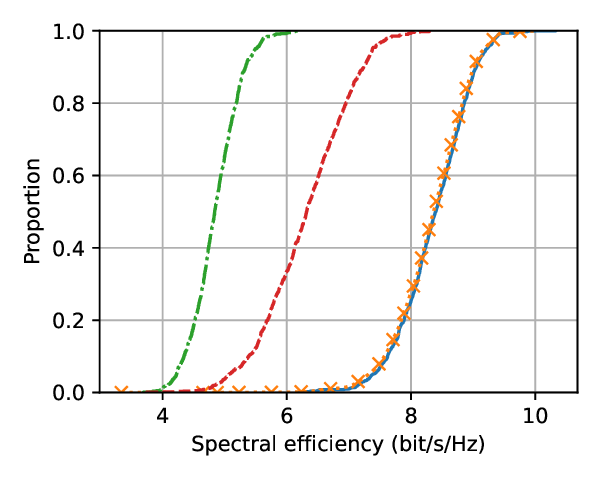}
}
\vskip 0cm
\subfloat[Rural NLoS: 96 APs, 36 UEs\label{fig:se_nlos_rural_96_36}]{
  \centering \includegraphics[width=0.48\textwidth, trim={0pt 10pt 10pt 10pt}, clip]{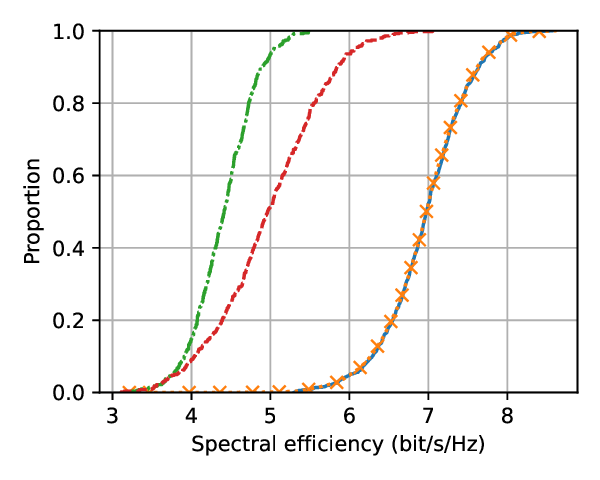}
}
\subfloat[Rural NLoS: 32 APs, 16 UEs\label{fig:se_nlos_rural_32_16}]{
  \centering \includegraphics[width=0.48\textwidth, trim={0pt 10pt 10pt 10pt}, clip]{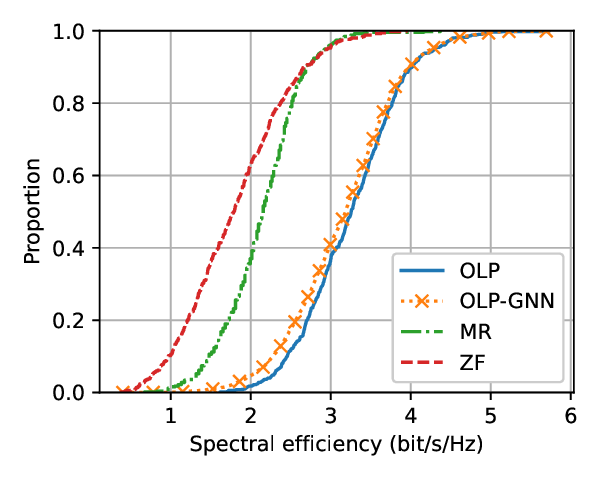}
}
\caption{Cumulative distribution functions of the downlink SE for MR, ZF, OLP and OLP-GNN for different environments and graph sizes.}
\label{fig:se_cdf}
\end{figure}

Figure~\ref{fig:se_cdf} shows the spectral efficiency of MR, ZF, OLP and OLP-GNN on four different scenarios. We first present the performance on scenarios with 96 APs and 36 UEs. These scenarios are relevant to evaluate the generalization of our solution since their graphs are bigger than the ones used for training, which have at most 64 APs and 18 UEs.
For the LoS environment in Fig.~\ref{fig:se_los_96_36}, OLP-GNN approximates the optimal with $1.30\%$ loss at median and a 95\%-likely SE loss of $2.52\%$.
Moreover, it outperforms MR and ZF precodings by respectively $63\%$ and $17\%$ at median.
We obtain similar results for the urban NLoS and rural NLoS scenarios in Fig.~\ref{fig:se_nlos_96_36} and~\ref{fig:se_nlos_rural_96_36}. In both cases, OLP-GNN has less than $0.5\%$ loss at median and $1\%$ loss at the 5-th percentile.
It significantly outperforms the baseline ZF by $32\%$ in the urban scenario and by $40\%$ in the rural scenario.

In Table~\ref{tab:gnn_vs_socp_time}, we observe 95\%-likely SE losses of at most $4\%$ for LoS datasets, $6\%$ for urban NLoS and $12\%$ for rural NLoS scenarios. 
Due to generalization error,
the bottom 5-th percentile is degraded in rural NLoS scenarios compared to their urban NLoS and LoS counterparts. Nonetheless, the median performance loss remains lower than $3\%$ in all environments.
Furthermore, even in the worst case scenario shown in Fig.~\ref{fig:se_nlos_rural_32_16}, our solution improves both median SE and 95\%-likely SE over MR by around $50\%$. It also outperforms ZF by respectively $78\%$ and $162\%$, at median and 5-th percentile respectively.

In NLoS scenarios with $24$ or $32$ APs, we see the performance degrading when the number of users $K$ increases. This is partly due to the difficulty of approximating matrix $\bA$ when the problem has lesser degrees of freedom and the UEs suffer greater interference. In this case, the diagonal elements of $\bA$ vary considerably, with some values being a magnitude of order higher than the others. Thus, the model trained in this study is best suited for "massive MIMO" systems where there are sufficiently more transmitting antennas $M$ than receivers $K$.

\begin{figure}
    \centering
    \includegraphics[width=0.95\columnwidth, trim={0 2pt 0 30pt}, clip]{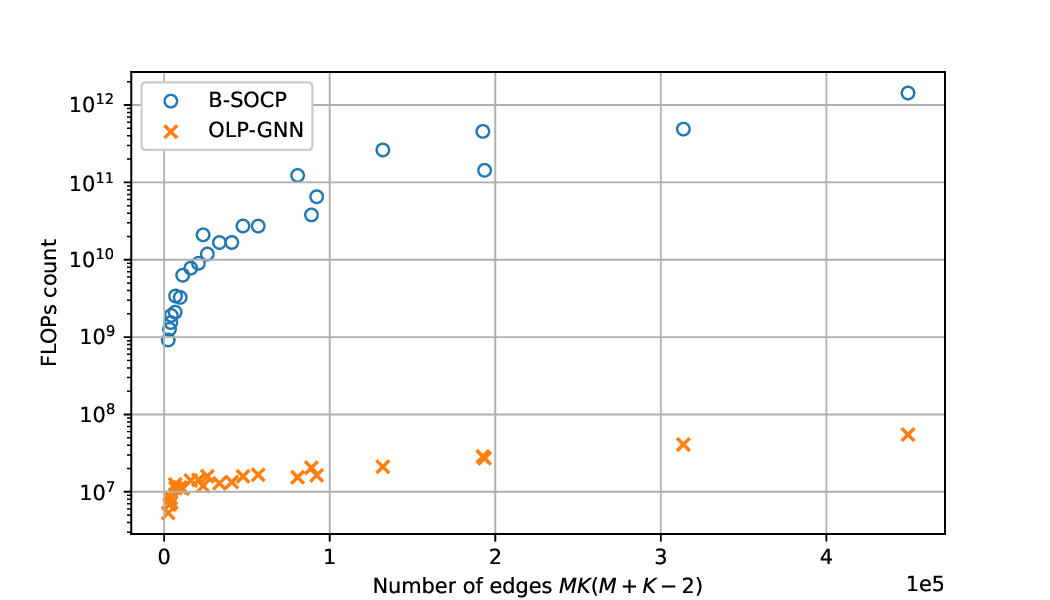}
    \caption{Number of FLOPs versus the number of edges for B-SOCP and OLP-GNN. Each point represents one rural NLoS scenario.}
    \label{fig:flops_count}
\end{figure}

\subsection{Computational Complexity and Runtime}

The asymptotic time complexity of OLP-GNN is $O\!\left(MK(M+K)\right)$. This can be derived by noting that the update rule~\eqref{eq:f_def} aggregates neighboring features with a complexity proportional to the number of edges.
The SOCP solver in MOSEK is based on primal-dual interior point method which has asymptotic complexity $O\!\left( n^{3.5}\log(\xi^{-1})\right)$~\cite{wright1997primal}, where $n=M+K$ in our problem and $\xi$ is the duality gap at termination. By applying bisection search on top of SOCP with a precision $\epsilon$, we deduce that B-SOCP runs in $O\!\left( (M+K)^{3.5}\log(\xi^{-1})\log(\epsilon^{-1})\right)$. In comparison, OLP-GNN has lower asymptotic complexity and does not have an iteration-complexity depending on a precision hyperparameter, $\xi$ or $\epsilon$.

In practice, we evaluate the complexity of our algorithms with the number of floating point operations (FLOPs). The FLOPs of our solution and the B-SOCP method are given in Fig.~\ref{fig:flops_count}. These FLOPs are obtained with PyPAPI (a tool to access low-level hardware performance counters) on an Intel Core i9-10980XE CPU. We see that \mbox{B-SOCP} requires respectively \num{5.7e2}, \num{4.0e3} and \num{1.2e4} times more FLOPs than OLP-GNN for $(M,K)$ $=$ $(32, 9)$, $(64, 18)$ and $(96, 27)$. As a consequence, OLP-GNN is several order of magnitude faster than B-SOCP.

We also measure the runtimes of OLP-GNN on a NVIDIA RTX A4000 GPU. Each dataset is repeated 10 times to obtain the runtime statistics in Table~\ref{tab:gnn_vs_socp_time}. These runtimes take into account preprocessing ($0.31$-$0.43$ms)
, OLP-GNN inference and postprocessing ($0.17$-$0.18$ms). For $24$ and $32$ APs, and up to $16$ UEs, the average runtimes are under 1ms. In the larger scenarios with up to 96 APs and 18 UEs, the runtimes are under 2ms. In all cases, the standard deviations (std) are lower than $0.01$ms, which indicates that these runtimes are consistently within the 1 to 2 millisecond requirement stated in the introduction.
This shows that OLP-GNN is implementable in practice with some limitations on the system size. Moreover, dedicated hardware and code optimization could further reduce the runtimes.

\section{Conclusions}
In this paper, we apply a graph neural network to the downlink max-min precoding problem in CFmMIMO. Our solution, named OLP-GNN, approximates the optimal linear precoder with several orders of magnitude faster runtimes than the state-of-the-art, making it feasible for real deployment for the first time. Indeed, the runtimes remain under 1ms for up to 32 APs and 16 UEs, and under 2ms for up to 96 APs and 18 UEs.

The characteristics of communication channels between transmitters and receivers can vary greatly.
We evaluate our trained model on both LoS/NLoS and urban/rural use-cases, demonstrating its generalizability to different environments and system sizes.
Simulations show that the median spectral efficiencies achieved by OLP-GNN are less than $3\%$ away from optimal on all scenarios.
 
Reducing further the time complexity of our solution would enable its execution on less powerful and costly hardware. In the current work, OLP-GNN takes as input $\bG^\dag$ which must be computed beforehand. Computing such a pseudo-inverse using classical numerical methods causes some overhead on the preprocessing time. It is therefore desirable to develop an end-to-end GNN without the above overheads.
We note that one way to speed up the inference time may be to apply GNN-specific quantization methods~\cite{tailor2020degreequant}.

\bibliographystyle{splncs04}
\bibliography{IEEEabrv, reference}

\end{document}